\documentclass{xray_symp}
\usepackage{graphics}

\begin{document}
\title{The X-Ray Luminosity Function of Active Galactic Nuclei}

\author{M. Schmidt \inst{1} \and R. Giacconi \inst{2} \and
        G. Hasinger \inst{3} \and J. Tr\"umper \inst{4} \and 
        G. Zamorani \inst{5,6}  }
 
\institute{California Institute of Technology, Pasadena, CA 91125, USA
\and European Southern Observatory, Karl-Schwarzschild-Str. 1,
     85748 Garching bei M\"unchen, Germany
\and Astrophysikalisches Institut Potsdam, An der Sternwarte 16,
     14482 Potsdam, Germany
\and Max-Planck-Institut f\"ur extraterrestrische Physik,
     Karl-Schwarzschild-Str. 2, 85740 Garching bei M\"unchen, Germany
\and Osservatorio Astronomico, Via Zamboni 33, 40126 Bologna, Italy
\and Istituto di Radioastronomia del CNR, via Gobetti 101,
     I-40129, Bologna, Italy}

\maketitle

\begin{abstract}

We derive an X-ray luminosity function for active galactic nuclei
(AGN) that accounts for the X-ray source counts in the 0.5--2.0
and 2--10 keV energy ranges, the redshift distribution of AGNs in the ROSAT
Deep Survey (RDS), as well as the X-ray background (XRB) from
1--10 keV. We emphasize the role of X-ray absorption, which has a 
large effect on the faint end of the 2--10 keV source counts, as
well as on the integrated X-ray background.
 
\end{abstract}

\section{Introduction}

This contribution for Joachim Tr\"umper's birthday marks his iniative and
involvement in the ROSAT Deep Survey (RDS). Little was known about the
luminosity function of X-ray sources when we started the planning meetings
for the survey some 13 years ago. Now that the main results of the RDS have
been published (Hasinger et al. 1998, Schmidt et al. 1998), this celebration
provides an excellent opportunity to review what we have learned about
the luminosity function of X-ray sources. Since a large fraction of these 
are active galactic nuclei (AGN), we will concentrate
on the AGN luminosity function.

Some of the earliest medium deep ROSAT fields were used by Boyle et al. (1993)
in combination with the Einstein Medium Sensitivity Survey (EMSS; Gioia
et al. 1990) to derive the AGN luminosity function and its evolution.
They concluded that the data are consistent with pure luminosity
evolution proportional to $(1+z)^{2.5}$ for $z < 2$ (for $q_o = 0.5$) 
and that the
integrated contribution of AGNs to the 2 keV X-ray background is 35\%. 
The effect of the uncertainty in the relative calibration of EMSS and
ROSAT (Page et al. 1996) can now be avoided by deriving the luminosity
function solely from ROSAT data. Hasinger (1998) and Miyaji et al. (1999)
have discussed all ROSAT survey material available and conclude that the 
luminosity functions of AGNs at different redshifts are incompatible 
with luminosity evolution but instead are consistent with density 
evolution. This result is important for our understanding of the
XRB, for, as we shall see below, under density evolution the entire
XRB can easily be explained as the integrated effect of AGNs.
 
The spectral difference between the X-ray background (XRB) 
and the typical cosmic X-ray 
source (AGN) is a consequence of the internal absorption in AGNs, which 
makes the spectrum of the more absorbed AGNs effectively very hard
(Setti and Woltjer 1989). Global analyses of the luminosity function 
of AGNs, such as those of Comastri et al. (1995), and Miyaji et al. 
(1998a) include the effects of absorption in accounting for the source 
counts $N(>S)$ and the XRB over a large energy range. The Comastri et al.
models adopt the Boyle et al. (1993) luminosity function and luminosity
evolution, while the Miyaji et al. models use a luminosity function
involving density evolution.

Our derivation of the AGN luminosity function will be more detailed but
less global than the above studies. We will only consider data from the 
brightest and deepest surveys at both 0.5--2.0 and 2--10 keV, as well
as the XRB in the 1--10 keV range.
We use a Hubble constant $H_o = 50$ km sec$^{-1}$ Mpc$^{-1}$ and a
deceleration parameter $q_o = 0.5$ throughout this paper.

\section {Methodology}

We will follow the method of Schmidt and Green (1986) in deriving the
luminosity function. We start from a well defined X-ray sample with 
optical identifications and redshifts (the {\it generating} sample).
For each of the sources in the generating sample, we derive the
maximum redshift $z_{max}$ at which it would be observed at the sample limit.
Since some of the AGNs are very nearby, we take into account the
local density enhancement (Santiago and Strauss 1992); specifically we
assume that the enhancement is a factor of 3 for $z<0.003$, and that it
is $4.5 - 500z$ for $0.003<z<0.007$.
We use an assumed law for the density evolution (to be iterated) in
evaluating the density-weighted volume $V_{max}$ over which each source is 
observable within the limits of the generating sample. The local luminosity
function is the sum of the delta functions $1/V_{max}$ of all sources 
in the sample. 

In deriving the redshift $z_{max}$, the spectral energy distribution of the
source plays an important role. This may be characterized by one or more
spectral indices, and absorption given by an effective hydrogen column
(cf. Morrison and McCammon 1983).
We assume that the energy spectral index of AGNs is $-1.3$ below
1 keV, and $-0.7$ above 1 keV (Schartel et al. 1997). Information about
the distribution of absorptions in complete AGN samples is very rare:
the HEAO1-A2 survey of X-ray sources at 2--10 keV (Piccinotti
et al. 1982) is the only complete sample for which hydrogen columns 
of the AGNs are known (Schartel et al. 1997) at the present time. 
The survey contains 30 AGNs to a limiting flux of $2.7\,10^{-11}$ cgs 
at 2--10 keV over an area of 27,020 square degrees. We use the Piccinotti
survey as the generating sample, thus ensuring that the observed 
properties of its sources, including the individual
absorptions, are precisely incorporated in the derivation of the
luminosity function.

Once the luminosity fuction is derived, we can predict the source counts,
redshift distributions and the integrated contribution of the AGNs to the XRB.
We then iterate the free parameters characterizing the density evolution
by fitting to observed source counts from other complete samples and/or 
the XRB, as described in the next section. 

\section {Results}

Since pure density evolution of the luminosity function leads to a
severe overestimate of the XRB (see below), we assume that the density 
evolution depends on the X-ray 2-10 keV luminosity (HX), such that the
co-moving density of sources varies with redshift as $(1+z)^k$ where

       $k = k_o$     
\hskip1.80in    for $ HX \gid HX_o$

       $k = k_o + k_1$ (log $HX$ - log $HX_o$) 
\hskip0.21in    for $ HX < HX_o$

\begin{figure}
 \resizebox{\hsize}{!}{\includegraphics{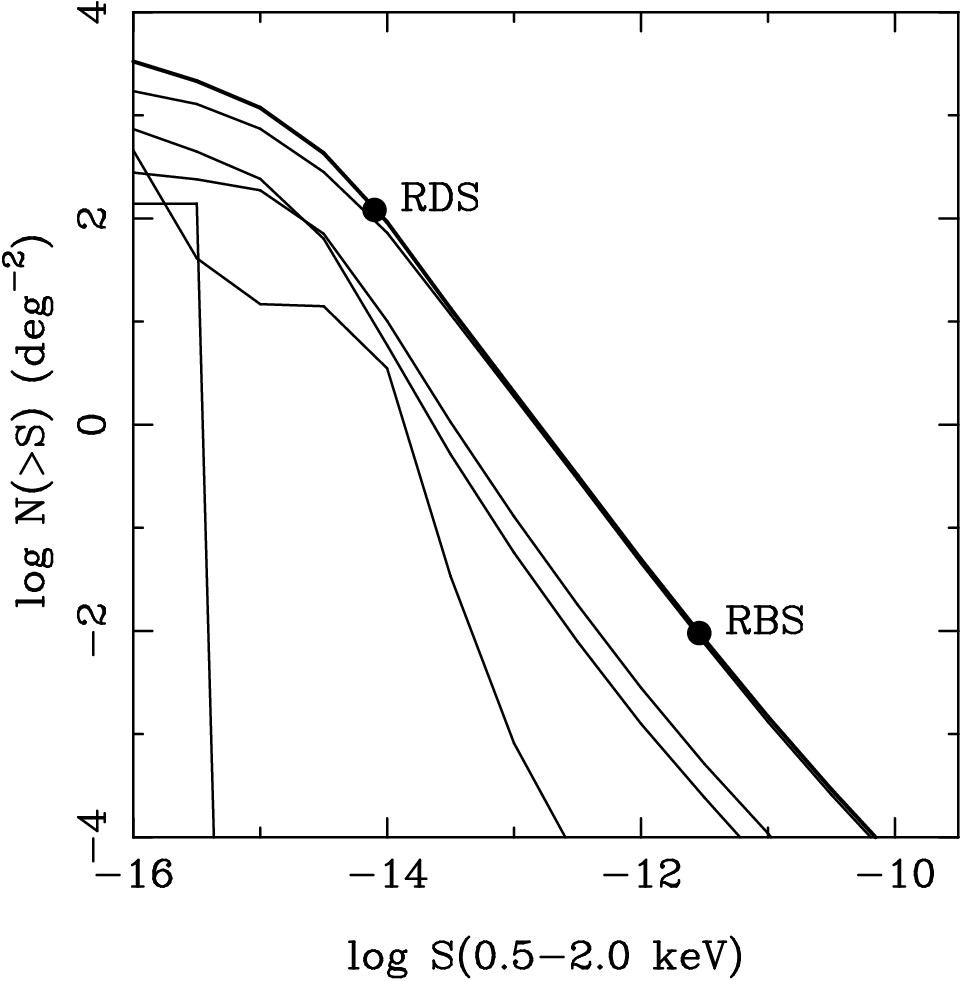}}
\caption[]{Predicted AGN source counts (thick curve) in the energy
range 0.5--2.0 keV. The thin curves represent the counts 
(from top to bottom, on the right side) for sources with columns 
$0$, $10^{21}$, $10^{22}$, $10^{23}$, and $10^{24}$ cm$^{-2}$.
The AGN counts in the RDS and the RBS are indicated (see text).}
\end{figure}

\noindent
At large redshifts, we adopt the evolution found for optically selected
quasars, viz. a freeze of the density increase for $z > 1.65$ (Hewett et al.
1993) and beyond $z = 2.7$ a decrease of the co-moving density
by a factor of 2.7 per unit redshift (Schmidt et al. 1995).

Our procedure is to set the free parameters 
characterizing the density evolution, $k_o$, $k_1$ and $HX_o$, such
that we fit the number of AGNs in the RDS, as well as the XRB in the
energy range 1-2 keV. The RDS covers 0.162 sq. deg. to a limit of
$1.1\,10^{-14}$ cgs and 0.136 sq. deg. to $0.55\,10^{-14}$ cgs. 
The sample contains 42 AGNs (Schmidt et al. 1998; source 36 has 
since been identified as an AGN at redshift 1.52; sources 14 and 84 have 
been tentatively identified with infrared objects in the K-band,
presumably reddened AGNs). For the XRB, we use the results of a
study by Miyaji et al. (1998b), recently updated by Miyaji
et al. (1998c). 

In iterating the free parameters for the density evolution, we first 
explore the case of pure density evolution, i.e., $k_1 = 0$.
In this case, the RDS number density is reproduced for $k_o = 4.75$,
but the predicted XRB at 1--2 keV for AGNs is twice the observed intensity.
We conclude that pure density evolution with $k_1 = 0$ is untenable.

Through trial and error, we find that both the RDS source count 
and the 1--2 keV XRB can be fitted with $k_o = 4.94$, $k_1 = 3.00$,
and log $HX_o = 44.00$. This is the model that we discuss in the remainder 
of this paper. It predicts an RDS redshift distribution in the redshift
intervals of 0--1, 1--2, 2--3, $>3$ of 12.3, 23.3, 6.1, and 0.4,
respectively,  where the 
observed numbers are 16, 20, 4, and 0, with 2 unknown. Since luminosities
and redshifts are strongly correlated in a flux-limited sample, the
excellent agreement shows that the rate of evolution for different
luminosities is essentially correct, confirming the luminosity dependence 
of the evolution to first order.

\begin{figure}
 \resizebox{\hsize}{!}{\includegraphics{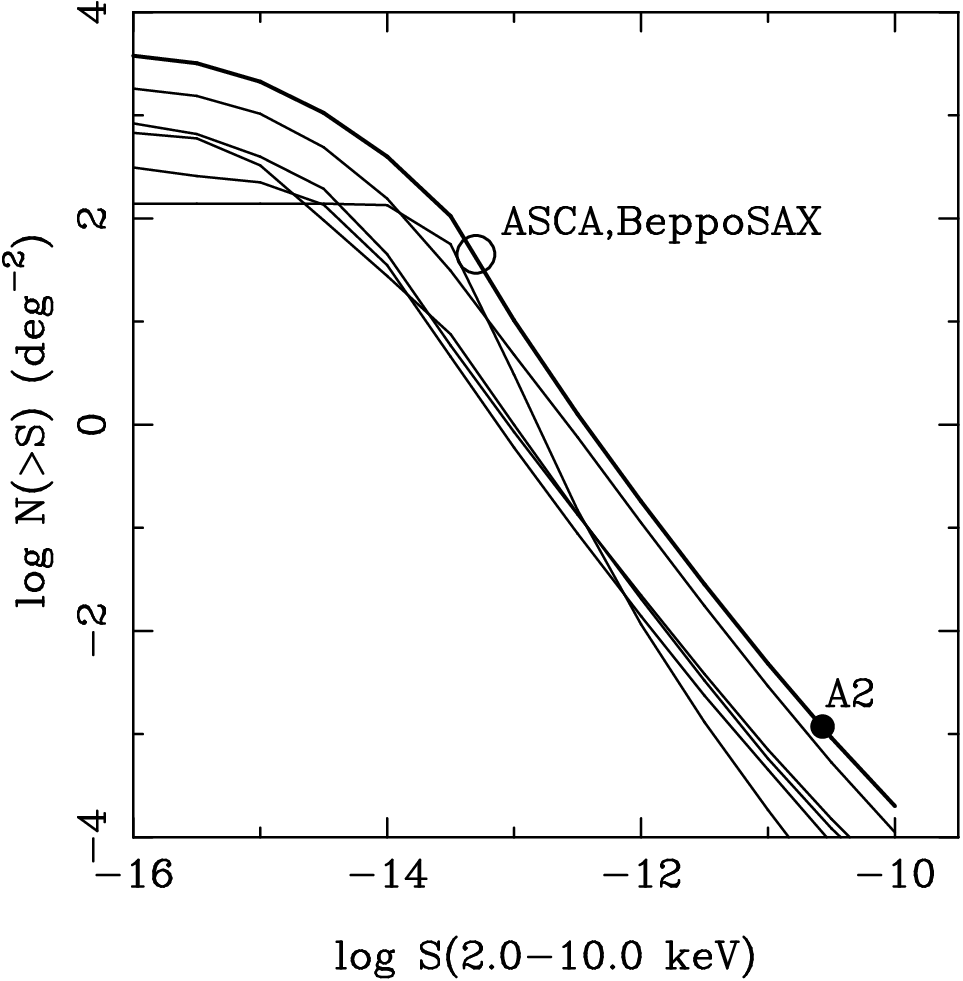}}
\caption[]{Predicted AGN source counts (thick curve) in the energy
range 2--10 keV. The thin curves represent the counts (from top to
bottom, both at the extreme left and right sides), for sources with 
columns $0$, $10^{22}$, $10^{23}$, $10^{21}$, and $10^{24}$ cm$^{-2}$.
The HEAO1-A2 AGN source count and a total source count representative of
the ASCA and BeppoSAX deep surveys (see text) are indicated.} 
\end{figure}

We test the model on the ROSAT Bright Survey (RBS, Schwope et al., 1998),
which is based on the ROSAT survey bright source catalogue 
(Voges et al. 1995). The RBS contains 194 AGNs with 
$0.5-2.0$ keV fluxes above $2.9\,10^{-12}$ cgs over an
area of 20,320 sq. deg. Our model predicts 193 sources. Considering the
small number of sources in our generating sample, the precise
agreement is fortuitous, but it is encouraging that there is no
discrepancy between these two large-area bright surveys in different
energy bands. Fig. 1 shows the predicted source counts at 0.5-2 keV, 
as well as the separate contributions from sources with different columns
 $N_H$. Further discussion of the $N_H$ distribution is given below.

Next, we turn to the source counts at 2--10 keV. At the bright end, the
AGN source count is provided by the Piccinotti survey, which is exactly
reflected in the model. At the faint end, we represent the results 
of deep surveys with ASCA (Cagnoni et al. 1998) and BeppoSAX (Giommi et 
al. 1998) by a representative source density of 45 deg$^{-2}$  
at $5\,10^{-14}$ cgs. Our model as presented so far yields only 
23 deg$^{-2}$.

This discrepancy is a direct consequence of an apparent incompatibility
between the 2--10 kev and the ROSAT 0.5--2.0 keV source counts. At a
source count of, say, $10\,{\rm deg}^{-2}$, the 2--10 keV flux from the ASCA
counts (Cagnoni et al. 1998) and the 0.5--2.0 keV ROSAT counts (Hasinger 
et al. 1998), if interpreted in terms of a single power law of the spectral
energy distribution, require a spectral index of $-0.5$. 
Typical observed spectral indices are $-0.7$ in the 2--10 keV band, 
and $-1.3$ in the 0.5--2 keV band (Schartel et al. 1997).
The spectral discrepancy suggests the existence of some sources with
a much harder spectrum or larger absorption than those of the
typical sources. The distribution of the effective hydrogen columns 
for the Piccinotti AGNs given by Schartel et al. (1997) is 18, 3, 5, and 4
for columns $0$, $10^{21}$, $10^{22}$, and $10^{23}$ cm$^{-2}$, 
respectively. We now explore the
effect of adding one or two sources with $N_H = 10^{24}$ to this sample.
This {\it ad hoc} addition is statistically not unreasonable: the 
probability of missing two marked sources out of 32 is around 13\%.  

\begin{figure}
 \resizebox{\hsize}{!}{\includegraphics{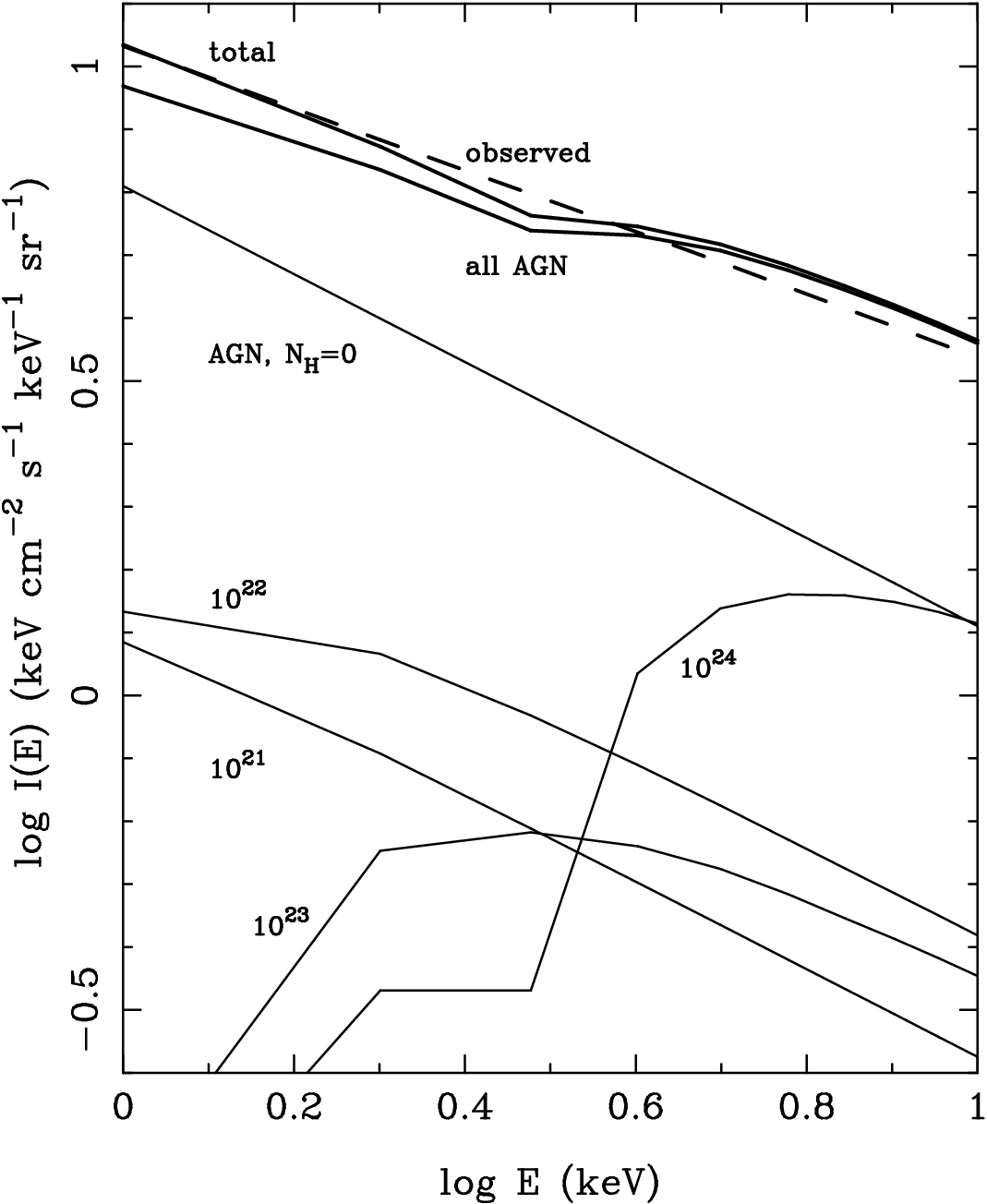}}
\caption[]{Predicted X-ray background from 1--10 keV. The contributions
from AGNs with different $N_H$ values are indicated, and their sum
is labeled ``all AGN''. The curve labeled ``total'' includes estimated 
contributions from galaxies and clusters of galaxies. The dashed curve
represents the observed XRB according to Miyaji et al. (1998b,c).}
\end{figure}

Specifically, we have added one {\it hypothetical} source to the Piccinotti
sample, with a flux of $3.2\,10^{-11}$ cgs, 
a redshift of 0.028 and log $N_H = 24.0$. The effect of this
additional source on the 0.5--2.0 keV source counts is negligible, as
shown by Fig. 1, as is its effect on the 1--2 keV XRB. Therefore, we can
leave the evolution parameters $k_o$, $k_1$, and HX$_o$ unchanged.
In contrast, the effect on the predicted 2--10 keV source counts is 
dramatic: at $5\,10^{-14}$ cgs the addition of this one source doubles 
the predicted counts, which now agree with the observed counts. Fig. 2 shows
that the log $N_H = 24$ fraction of the counts rises with decreasing flux,
reaches a maximum between $10^{-13}$ and $10^{-14}$ cgs 
and then declines again. This
complex behavior is a consequence of the spectral properties of a
heavily absorbed source moving through the 2--10 keV band as its redshift
increases. The redshift of 0.028 for the hypothetical 
source was chosen to maximize the effect on the source counts. 
For redshifts of 0.024 and 0.064, respectively, the effect on the 
2--10 keV source counts would be half as large, so two such sources 
would be required for the same effect. There are 10 sources in the
Piccinotti sample with redshifts in the range 0.024--0.064.

The X-ray background is illustrated in Fig. 3. The curve labeled "all AGN"
is the sum of the five components with different $N_H$ values. The "total"
curve includes an estimate of the background expected from galaxies, and
clusters of galaxies. The good fit to the observed XRB (Miyaji et al. 
1998b,c) at 1-2 keV is the result of our choice of $k_o$ and $k_1$, see 
above. The log $N_H = 24$ component, generated by our hypothetical source, 
has a substantial effect on the predicted background above 4 keV, 
resulting in good agreement with the observed background at higher energies.

\section {Discussion}

We have succeeded in deriving a model of the AGN luminosity function and its
evolution that:

\begin{enumerate}
\item at 0.5--2.0 keV reproduces the number of AGNs in the RBS and the RDS,
   as well as the redshift distribution in the RDS;
\item at 2--10 keV reproduces the number and redshift
   distribution of AGNs in the HEAO1-A2 survey, as well as the source
   counts in the deep ASCA and BeppoSax surveys
\item reproduces the observed XRB from 1--10 keV.
\end{enumerate}

Essential ingredients of our derivation of the luminosity function are 
({\it a}) the luminosity-dependent density evolution, 
with a cutoff at high redshift, and ({\it b}) evaluation of the 
effect of a realistic distribution of absorption columns. 
The success of our model analysis supports the proposals by 
Setti and Woltjer (1989), Madau et al.(1994), and others that the spectral 
difference between the XRB and the typical cosmic X-ray source (AGN) 
is a consequence of the internal absorption in AGNs, which makes the 
spectrum of the more absorbed AGNs effectively very hard. It appears
from our model that the high source counts at 2--10 keV compared to
those at 0.5--2.0 keV are caused by the same effect. 


The small number of AGNs in the HEAO1-A2 survey inherently limits the 
statistical accuracy of the model we have discussed here. 
The model predictions for the number of sources expected in the RBS, the RDS,
etc. generally have a corresponding statistical error of around 25\%.
At energies above 2 keV, the lack of statistics about the fraction of
sources with log $N_H = 24$ is a major source of uncertainty, as we
discussed in Sec. 3. Our attempt to represent this component by a
single hypothetical source in the Piccinotti sample that we used as a
generating sample, allowed us to illustrate the effect of these sources.
Statistical information on the frequency of high absorption columns
will be required to confirm or refine luminosity function models such
as presented in this paper. Recent BeppoSAX observations of a sample of
low luminosity AGNs selected on the basis of [O III] fluxes, assumed to
be an isotropic luminosity indicator, show that the number of highly
absorbed sources is substantially higher than previously assumed on the
basis of existing 2--10 keV data (Maiolino et al. 1998).
Once the statistical data on absorption
columns are available, the question of how to take into account 
the wide variety of absorption spectra seen in AGN spectra 
(cf. Reichert et al. 1985), which has been ignored in this paper, 
should also be addressed.

Further systematic surveys are desired, especially at higher energies,
preferably beyond 10 keV, both shallow over a large sky area
as well as deeper surveys over smaller areas. Optical identifications 
and redshifts for all sources, or well defined subsamples, are needed, 
as well as X-ray spectra of sufficiently large samples so that 
the distribution of absorption columns to $> 10^{24} cm^{-2}$ 
can be determined. 

\vskip 0.4cm

\begin{acknowledgements}
The ROSAT project is supported by the Bundesministerium
f\"ur Forschung und Technologie (BMFT), by the National
Aeronautics and Space Administration (NASA), and the Science
and Engineering Research Council (SERC). The W. M. Keck Observatory
is operated as a scientific partnership between the California
Institute of Technology, the University of California, and the
National Aeronautics and Space Administration. It was made possible
by the generous financial support of the W. M. Keck Foundation.
M.S. thanks the Alexander von Humboldt-Stiftung for a Humboldt Research
Award for Senior U.S. Scientists in 1990-91; and the directors of the
Max-Planck-Institut f\"ur extraterrestrische Physik in Garching and of
the Astrophysikalisches Institut Potsdam for their hospitality.
We thank T. Miyaji for results on the X-ray background before publication.
This work was supported in part by NASA grants NAG5-1531 (M.S.), 
NAG8-794, NAG5-1649, and NAGW-2508 (R.G.). G.H. acknowledges the DARA grant 
FKZ 50 OR 9403 5. G.Z. acknowledges partial support by the Italian 
Space Agency (ASI) under contracts 95-RS-152 and ARS-96-70.
We gratefully acknowledge the permission by the Springer Verlag
to use their A\&A \LaTeX{} document class macro.
\end{acknowledgements}

\end{document}